\documentclass{elsart}
\newcommand{\be}{\begin{eqnarray}}
\newcommand{\ee}{\end{eqnarray}}
\usepackage{amsmath}
\usepackage{amssymb}
\usepackage{epsfig}

\newcommand{\dis}{\displaystyle}
 
\begin{document}
\hspace{9.8 cm}FZJ--IKP(TH)--2003--09
\begin{frontmatter}
\title{How to extract the $\Lambda N$ scattering length from 
production reactions}
 
\author{A. Gasparyan$^{1,2}$, J. Haidenbauer$^1$, C. Hanhart$^1$, and J. Speth$^1$}

{\small $^1$Institut f\"{u}r Kernphysik, Forschungszentrum J\"{u}lich GmbH,}\\ 
{\small D--52425 J\"{u}lich, Germany} \\
{\small $^2$Institute of Theoretical and Experimental Physics,}\\
{\small 117259, B. Cheremushkinskaya 25, Moscow, Russia}\\
 
\begin{abstract}
  A dispersion integral is derived that allows one to relate directly (spin dependent)
  $\Lambda N$ invariant mass spectra, measured in a large-momentum transfer
  reaction such as $pp\to K^+p\Lambda$ or $\gamma d\to K^+n\Lambda$, to the
  scattering length for elastic $\Lambda N$ scattering. The involved
  systematic uncertainties are estimated to be smaller than 0.3 fm. This
  estimate is confirmed by comparing results of the proposed formalism with
  those of microscopic model calculations.  We also show, for the specific
  reaction $pp\to K^+\Lambda p$, how polarization observables can be used to
  separate the two spin states of the $\Lambda N$ system.
\end{abstract}

\end{frontmatter}

\section{Introduction}
Nucleons and hyperons form a flavor--$SU(3)$ octet. The underlying $SU(3)$
symmetry is clearly broken, as is evidenced by the mass splittings within
the members of the octet. This symmetry breaking can be well accounted for 
with relations such as the Gell-Mann--Okubo formula. 
The interesting question is, however, whether there is also a dynamical
breaking of the $SU(3)$ symmetry - besides these obvious ``kinematic''
effects. 

The nucleon--nucleon ($NN$) interaction has been studied in
great detail for many decades and we have a good understanding of this system
at energies even beyond the pion production threshold.  On the other hand,
very little is known about the dynamics involving the other members of the
octet. 
Because of that it has become common practice in studies of hyperon physics
to assume {\it a priori} $SU(3)$ flavor symmetry. Specifically, in meson exchange models
of the hyperon-nucleon ($YN$) interaction such symmetry requirements provide
relations between coupling constants of mesons of a given multiplet to
the baryon current, which greatly reduce the number of free model parameters.
Then coupling constants at the strange vertices are connected to 
nucleon-nucleon-meson coupling constants, which in turn are constrained
by the wealth of empirical information on $NN$ scattering  
\cite{dover,maessen,juel,juel1,rijken,jueln}. 
The scarce and not very accurate data set available so far for elastic 
$YN$ scattering \cite{data1,data2,data3} seems to be
indeed consistent with the assumption of $SU(3)$ symmetry. 
Unfortunately, the short lifetime
of the hyperons hinders high precision scattering experiments at low energies 
and, therefore, has so far precluded a more thorough test of the validity of 
$SU(3)$ flavor symmetry. 

The poor status of our information on the $YN$ interaction is
most obviously reflected in the present knowledge of the $\Lambda N$ 
scattering lengths. Attempts in the 1960's to pin down the low
energy parameters for the $S$-waves led to results that were afflicted by rather 
large uncertainties \cite{data1,data1b}. In Ref. \cite{data1b}
the following values are given for the singlet scattering length $a_s$ and the
triplet scattering length $a_t$
\begin{equation}
a_s=-1.8\left\{{+2.3 \atop -4.2}\right. \ \mbox{fm and } 
a_t=-1.6\left\{{+1.1 \atop -0.8}\right. \ \mbox{fm} ,
\end{equation}
where the errors are strongly correlated. The situation of the corresponding
effective ranges is even worse: for both spin states values between 0 and 16 fm
are allowed by the data.
Later, the application of microscopic models 
for the extrapolation of the data to the threshold, was hardly more successful. 
For example, in 
Ref. \cite{rijken} one can find six different models that equally well
describe the available data but whose ($S$-wave) scattering lengths 
range from -0.7 to -2.6 fm in the singlet channel and from -1.7 to -2.15 fm 
in the triplet channel.

A natural alternative to scattering experiments are studies of production
reactions.  In Ref. \cite{ben1} it was suggested to use the reaction $K^-d\to
n\Lambda \gamma$, where the initial state is in an atomic bound state, to
determine the $\Lambda N$ scattering lengths. From the experimental side so
far, a feasibility study was performed which demonstrated that a separation of
background and signal is possible \cite{kdexp}.  The reaction $K^-d\to
n\Lambda \gamma$ was studied theoretically in more detail in Refs.
\cite{akh,workman,ben2}. The main results especially of the last work are that
it is indeed possible to use the radiative $K^-$ capture to extract the
$\Lambda N$ scattering lenghts and that polarization observables could be used
to disentangle the different spin states.  In that paper it was also shown,
however, that to some extend the extraction is sensitive to the short range
behavior of the $YN$ interaction.  The reaction $K^-d\to \pi^-p\Lambda$ was
analyzed in Ref. \cite{tan}, leading to a scattering length of $-2\pm 0.5$ fm
via fitting the invariant mass distribution to an effective range
expansion---the author argued that this value is to be interpreted as the spin
triplet scattering length.  It is difficult to estimate the theoretical
uncertainty and the error given is that of the experiment only.

In the present paper we argue that large-momentum transfer reactions such as
$pp\to K^+p\Lambda$ \cite{C11_1,Bilger,jan} or $\gamma d\to K^+n\Lambda$
\cite{Renard,Mecking,Adel,Li,Yama} might be the best candidates for extracting
informations about the $\Lambda N$ scattering lengths. In reactions with large
momentum transfer the production process is necessarily of short-ranged
nature. As a consequence the results are basically insensitive to details of
the production mechanism and therefore a reliable error estimation is
possible. In Ref. \cite{jan} the reaction $pp\to K^+p\Lambda$ at low
excess energies was already used to determine the low energy parameters of the
$\Lambda N$ interaction. 
The authors extracted an average value of 
$-2\pm 0.2$ fm for the $\Lambda N$ scattering length in an analysis that 
utilizes the effective range expansion. But also this work has some
drawbacks. First, again the given error is statistical only. More
serious, however, is the use of the effective range expansion. As we will
show below this is only appropriate for systems in which the scattering 
length is significantly larger than the effective range. In addition,
using this procedure one 
encounters strong correlations between the effective range parameters 
$a$ and $r$ that can only be disentangled by including other data, 
e.g. $\Lambda N$ elastic cross sections, into the analysis \cite{jan}. 

In this manuscript we propose a method that allows 
extraction of the $YN$ scattering lengths from the production data
directly.  In particular, we derive an integral representation
for the $\Lambda N$ scattering lengths in terms of a differential
cross section of reactions with large momentum transfer 
such as $pp\to K^+p\Lambda$ or $\gamma d\to K^+n\Lambda$. It reads
\begin{eqnarray}
\nonumber
a_S&=&\lim_{{m}^2\to m_0^2}\frac1{2\pi}\left(\frac{m_\Lambda+m_N}
{\sqrt{m_\Lambda m_N}}\right){\bf P}
\int_{m_0^2}^{m_{max}^2}dm' \, ^2
\sqrt{\frac{m_{max}^2-{m}^2}{m_{max}^2-m' \, ^2}}\\
& & \qquad \qquad \times \ 
\frac1{\sqrt{m' \, ^2-m_0^2} \ (m' \, ^2-{m}^2)}
\log{\left\{\frac{1}{p'}\left(\frac{d^2\sigma_S}{dm' \, ^2dt}\right)\right\}}
\ ,
\label{final}
\end{eqnarray}
where $\sigma_S$ denotes the spin cross section for the production of a
$\Lambda$--nucleon pair with invariant mass $m' \, ^2$---corresponding to a
relative momentum $p'$---and total spin $S$. In addition $t=(p_1-p_{K^+})^2$, 
with $p_1$
being the beam momentum, $m_0^2=(m_\Lambda+m_N)^2$, where $m_\Lambda$ ($m_N$)
denotes the mass of the Lambda hyperon (nucleon), and $m_{max}$ is some
suitably chosen cutoff in the mass integration.  We will argue below that it
is sufficient to include relative energies of the final $\Lambda N$ system of
at most 40 MeV in the range of integration to get accurate results.  $\bf P$
denotes that the principal value of the integral is to be used and the limit
has to be taken from above.  This formula should enable determination of the
scattering lengths to a theoretical uncertainty of at most 0.3 fm.  Note that,
due to recent progress in accelerator technology, the differential cross
section required can be measured to high accuracy even 
for small $\Lambda N$ energies.

The method we propose is applicable to all large momentum transfer reactions, 
as long as the final $\Lambda N$ system is dominated by a single partial wave. 
Evidently, the number of contributing partial waves is already strongly reduced 
by the kinematical requirements specified above. To be concrete, we do not 
expect that $P$ or higher partial waves are of relevance for $\Lambda N$ 
energies below the mentioned 40 MeV. However, in an unpolarized measurement 
both the spin triplet ($^3S_1$) as well as the spin singlet ($^1S_0$) final state 
can contribute with a priory unknown relative weight. 
Fortunately polarization observables allow to
disentangle model independently the different spin states. In appendix \ref{obs} we 
demonstrate this for the specific reaction $pp\to p\Lambda K^+$.

\section{Formalism}

\subsection{The Enhancement Factor}

A standard method to calculate the effect of a particular final state
interaction is that of dispersion relations \cite{watson}.
The production amplitude $A$ depends on the total energy squared 
$s=(p_1+p_2)^2$, the invariant mass squared of the
outgoing $\Lambda p$ system $m^2=(p_N+p_\Lambda)^2$ and the
momentum transfer $t$ defined above.
The dispersion relation for $A$ at fixed $s$ and $t$ and a specific
partial wave $S$ 
takes the form
\begin{eqnarray}
A_S(s,t,m^2)=\frac1\pi\int_{-\infty}^{\tilde m\, ^2} \frac{D_S(s,t,m' \, ^2)}{m' \, ^2-m^2}dm' \, ^2
+\frac1\pi\int_{m_0^2}^\infty \frac{D_S(s,t,m' \, ^2)}{m' \, ^2-m^2}dm' \, ^2,
\label{dispers}
\end{eqnarray}
where $m^2=\tilde m\, ^2$ is the lefthand singularity 
closest to the physical region and 
\begin{eqnarray}
D_S(s,t,m^2) = \frac{1}{2i}(A_S(s,t,m^2+i0)-A_S(s,t,m^2-i0)).
\end{eqnarray}
Here $S$ stands for the set of quantum numbers that characterize the
production amplitude $A$ projected on the hyperon--nucleon partial waves.
Since we 
restrict ourselves to $s$--waves in the $YN$ system, 
$S$ corresponds directly to the total spin of the $YN$ system. 
In order to simplify the notation we will omit the spin index in the 
following. 
 
The integrals in Eq. (\ref{dispers}) receive contributions
from the various possible final-state interactions, namely $\Lambda K$,
$NK$ as well as $\Lambda N$.
The first two can be suppressed by choosing the reaction kinematics properly
and thus we may neglect them for the moment---we come back to their
possible influence below---to get, 
for $m^2>m_0^2$,
\begin{eqnarray}
D(s,t,m^2) = A(s,t,m^2)e^{-i\delta}\sin{\delta},
\label{drhc}
\end{eqnarray}
where $\delta$ is the elastic $\Lambda N$ ($^1S_0$ or $^3S_1$) scattering phase 
shift\footnote{
Obviously this expression holds only for those values of $m^2$ that are below
the first inelastic threshold. We will ignore this  here and come back to 
the role of the inelastic channels later.
}. 

The solution of Eq. \eqref{dispers} in the 
physical region becomes (see Refs. \cite{Muskhelishvili1953,Omnes1958,Frazer1959})
\begin{eqnarray}
A(s,t,m^2)=e^{\dis{u(m^2+i0)}}
\frac1\pi\int_{-\infty}^{\tilde m\, ^2} \frac{dm' \, ^2D(s,t,m' \, ^2)}{m' \, ^2-m^2}e^{\dis{-u(m' \, ^2)}},
\label{Frazer}
\end{eqnarray}
where, in the absence of bound states, 
\begin{eqnarray}
u(z)=\frac1\pi\int_{m_0^2}^\infty\frac{\delta(m' \, ^2)}{m' \, ^2-z}dm' \, ^2.
\label{udef}
\end{eqnarray}
In a large-momentum transfer reaction,
the only piece with a strong dependence on $m^2$ is given by the exponential 
factor in front of the integral in Eq. (\ref{Frazer}).
We may thus define
\begin{eqnarray}
A(s,t,m^2)=\exp\left[{\frac1\pi\int_{m_0^2}^\infty\frac{\delta(m' \, ^2)}{m' \, ^2-m^2-i0}dm' \, ^2}\right]
\Phi(s,t,m^2),
\label{dis0}
\end{eqnarray}
where
 $\Phi(s,t,m^2)$ is a slowly varying function of $m^2$.
Henceforth we suppress the dependence of the amplitude 
on $s$ and $t$ in our notation.
In the literature $A(m^2)$ is known as the enhancement factor \cite{watson}.

It is interesting to investigate the form of $A(m^2)$ for phase shifts
that are given by the first two terms in the effective range
expansion,
\begin{eqnarray}
p' \ {\rm ctg} (\delta(m^2)) = -1/a+(1/2)rp' \,^2 \ , 
\label{ere}
\end{eqnarray}
over the whole energy range. 
Here $p'$ is the
relative momentum of the final state particles under consideration
in their center of mass system. $A(m^2)$ can then be given in closed
form as \cite{book}
\begin{eqnarray}
A(m^2) = \frac{(p' \, ^2+\alpha^2)r/2}{-1/a+(r/2)p' \, ^2-ip'}\Phi(m^2) \ ,
\label{arform}
\end{eqnarray}
where $\alpha = 1/r(1+\sqrt{1-2r/a})$.
Note that for $a\gg r$, as is almost realized in the $^1S_0$ partial 
wave of the $pp$ system,
the energy dependence of $A(m^2)$ is given by $1/(1+iap')$ as long as $p' \ll 1/r$. 
This, however, is identical with the energy dependence of $pp$ elastic scattering 
(if we disregard the effect of the Coulomb interaction). 
Therefore one expects that, at least for
small kinetic energies, $pp$ elastic scattering and meson production
in $NN$ collisions with a $pp$ final state exhibit the same energy dependence \cite{watson,migdal}, 
which indeed was experimentally confirmed by the measurements of the reaction 
$pp\to pp\pi^0$ \cite{IU1} close to threshold.
 
The situation is different, however, for interactions where the scattering length
is of the same order as the effective range, because then the numerator of
Eq. \eqref{arform} introduces a non--negligible momentum dependence.
This observation suggests that a large theoretical uncertainty is to 
be assigned to the scattering length extracted in Ref. \cite{jan}.
If, furthermore, even higher order terms 
in the effective range expansion are necessary
to describe accurately the phase shifts, as might be the case, 
e.g., for the $YN$ interaction \cite{ben2}, no closed form
expression can be given anymore for $A(m^2)$. Thus one would have 
to evaluate numerically the integral given in Eq. (\ref{udef}) for 
specific models and then compare the result to the data. 
This procedure is not very transparent. Therefore we propose
another approach to be outlined in the next subsection. 
Starting from Eq. (\ref{dis0}) we demonstrate 
that the scattering length can be extracted from the data directly.  
Furthermore we show how -- at the same time -- an estimate of  
the theoretical uncertainty in the extraction can be obtained.

\begin{figure}[t]
\vspace{5cm}
\includegraphics{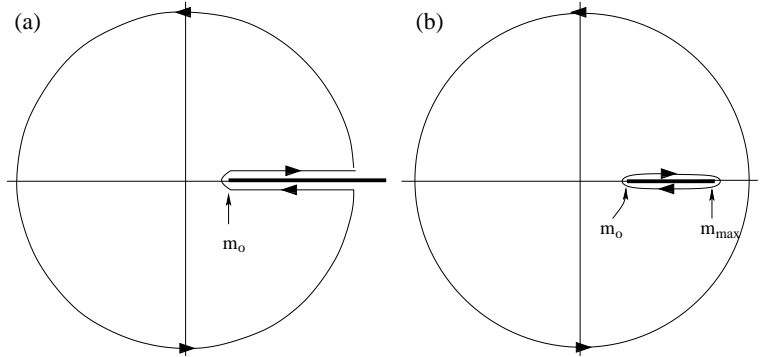}
\caption{The integration contours in the complex $m' \, ^2$ plane to be used to derive
Eqs. (\protect{\ref{dis2}}) (a) and (\protect{\ref{dis3}}) (b).
The thick lines indicate the branch cut singularities.}
\label{intcont} 
\end{figure}

\subsection{Extraction of the scattering length}

Our aim is to establish a formalism that facilitates the extraction of 
the $\Lambda N$ scattering length from experimental 
information on the invariant mass distribution. 
To do so we derive a dispersion relation for $|A(m^2)|^2$. 
We will use a method similar to 
the one utilized in Refs. \cite{Geshkenbein1969,Geshkenbein1998}.
First, notice that -- by construction -- the function 
\begin{eqnarray}
\frac{\log{\{A(m^2)/\Phi(m^2)\}}}{\sqrt{m^2-m_0^2}}
=\frac1{\sqrt{m^2-m_0^2}}\frac1\pi\int_{m_0^2}^\infty\frac{\delta(m' \, ^2)}{m' \, ^2-m^2-i0}dm' \, ^2
\label{dis1}
\end{eqnarray}
has no singularities on the physical sheet except for  
the cut from $m_0^2$ to infinity. In addition, its 
value below the cut is equal to the negative of the
complex conjugate of the corresponding value above the cut.
Therefore, from Cauchy's theorem we get
\begin{eqnarray}
\frac{\log{\{A(m^2)/\Phi(m^2)\}}}{\sqrt{m^2-m_0^2}}=
\frac1{2\pi i}\int_{m_0^2}^\infty\frac{\log{|A(m' \, ^2)/\Phi(m' \, ^2)|^2}}
{\sqrt{m' \, ^2-m_0^2} \, (m' \, ^2-m^2-i0)}dm' \, ^2 \ .
\label{dis2}
\end{eqnarray}
Taking the imaginary parts of Eqs. \eqref{dis1}
and \eqref{dis2} one gets
\begin{eqnarray}
\frac{\delta(m^2)}{\sqrt{m^2-m_0^2}}=-\frac1{2\pi}{\bf P}
\int_{m_0^2}^\infty\frac{\log{|A(m' \, ^2)/\Phi(m' \, ^2)|^2}}
{\sqrt{m' \, ^2-m_0^2} \, (m' \, ^2-m^2)}dm' \, ^2 \ . 
\label{inf_int}
\end{eqnarray}
Note that for small $\Lambda N$ kinetic energies we may write
 $m^2=m_0^2+\frac{\dis m_0 \ p'\, ^2}{\dis\mu}$, where
$\mu=\frac{\dis m_{\Lambda}m_N}{\dis m_{\Lambda}+m_N}$.
Thus, for small invariant masses, $\sqrt{m^2-m_0^2}\propto p'$, which clearly 
demonstrates that the left hand side of Eq. (\ref{inf_int}) converges to the
(negative of the) scattering length times $\sqrt{\mu / m_0}$ for $m^2\to m_0^2$.

The integral in Eq. \eqref{inf_int} depends on the behaviour of $A(m' \, ^2)$ for 
large $m' \, ^2$ which is not accessible experimentally. 
Therefore, in practice the integration in Eq. \eqref{inf_int} has to be restricted 
to a finite range. In fact, a truncation of the integrals is required anyway because in
our formalism we do not take into account explicitly (and we do not want to) 
the complexities arising from further right-hand cuts caused by the opening of the 
$\Sigma N$, $\pi \Lambda N$, etc., channels. 
Thus, let us go back to the definition of $A$ in Eq. (\ref{dis0}). A significant contribution 
of the integral there stems from large values of $m' \, ^2$. Those contributions depend 
only weakly on $m^2$, at least for $m^2$ values in the region close to threshold.
Therefore, they can be also absorbed into the function $\Phi$ 
defined in Eq. (\ref{dis0}). Consequently, we can write 
\begin{eqnarray}
A(m^2)=\exp\left[{\frac1\pi\int_{m_0^2}^{m_{max}^2}\frac{\delta(m' \, ^2)}{m' \, ^2-m^2-i0}dm' \, ^2}\right]
\tilde \Phi(m_{max}^2,m^2) \label{C_int} \ ,
\label{A_def} 
\end{eqnarray}
where $\tilde \Phi(m_{max}^2,m^2)=\Phi(m^2)\Phi_{m_{max}^2}(m^2)$ with
\begin{eqnarray}
 \Phi_{m_{max}^2}(m^2)=\exp\left[{\frac1\pi\int_{m_{max}^2}^{\infty}\frac{\delta(m' \, ^2)}{m' \, ^2-m^2-i0}dm' \, ^2}\right].
\label{dis3}
\end{eqnarray}
The quantity $m_{max}^2$ is to be chosen by physical arguments in
such a way that both $\Phi(m^2)$ and $\Phi_{m_{max}^2}(m^2)$ 
vary slowly over the interval $(m_0^2,m_{max}^2)$. In particular, it is crucial 
that it can be chosen to be smaller than the invariant mass corresponding to the
$\Sigma N$ threshold. Note also that, strictly speaking, Eq. \eqref{dis3} is not 
valid anymore in the presence of inelastic channels. 
Rather one should view Eq. \eqref{A_def} as the actual definition of 
$\tilde \Phi$. But we will still use Eq. \eqref{dis3} later to estimate the 
uncertainty of Eq. \eqref{final}.

We should mention 
that the integral in Eq. \eqref{C_int} contains 
an unphysical singularity of the type $\log{(m_{max}^2-m^2)}$ --  which is 
canceled by a corresponding singularity in $\tilde \Phi(m_{max}^2,m^2)$ -- 
but this again does not affect the region near threshold.

The method of reconstruction of $\delta(m^2)$  is similar to the
one used for the infinite range integration. The 
function 
\begin{eqnarray}
\nonumber
\frac{\log{\{A(m^2)/\tilde \Phi(m_{max}^2,m^2)\}}}{\sqrt{(m^2-m_0^2)(m_{max}^2-m^2)}}
&=&\frac1{\sqrt{(m^2-m_0^2)(m_{max}^2-m^2)}} \\
&\times & \
\frac1\pi\int_{m_0^2}^{m_{max}^2}\frac{\delta(m' \, ^2)}{m' \, ^2-m^2-i0}dm' \, ^2
\end{eqnarray}
has no singularities in the complex plane except
the cut from $m_0^2$ to $m_{max}^2$. Again, its value
below the cut is equal to the negative of the complex conjugate 
of its value above the cut.
Hence, applying Cauchy's theorem, one gets
\begin{eqnarray}
\nonumber
\frac{\log{\{A(m^2)/\tilde \Phi(m_{max}^2,m^2)\}}}{\sqrt{(m^2-m_0^2)(m_{max}^2-m^2)}}&=& \\
\nonumber
& & \!\!\!\!\!\!\!\!\!\!\!\!\!\!\!\!\!\!\!\!\!\!\!\!\!\!\!\!\!\!\!\!\!\!\!\!\!\!\!\!
\frac1{2\pi i}\int_{m_0^2}^{m_{max}^2}\frac{\log{|A(m' \, ^2)/\tilde \Phi(m_{max}^2,m' \, ^2)|^2}}
{\sqrt{(m' \, ^2-m_0^2)(m_{max}^2-m' \, ^2)}(m' \, ^2-m^2-i0)}dm' \, ^2,
\label{dis4}
\end{eqnarray}
and, accordingly, 
\begin{eqnarray}
\nonumber
&&\frac{\delta(m^2)}{\sqrt{m^2-m_0^2}}= \\
& & -\frac1{2\pi}{\bf P}
\int_{m_0^2}^{m_{max}^2}\frac{\log{|A(m' \, ^2)/\tilde \Phi(m_{max}^2,m' \, ^2)|^2}}
{\sqrt{m' \, ^2-m_0^2} \ (m' \, ^2-m^2)}\sqrt{\frac{m_{max}^2-m^2}{m_{max}^2-m' \, ^2}}dm' \, ^2.
\label{almostfinal}
\end{eqnarray}
It is important to stress that the principal value integral vanishes for a 
constant argument of the logarithm as long as $m_0^2 \le m^2 \le m_{max}^2$. 
Therefore, if the
function $\tilde \Phi$ depends only weakly on $m'\, ^2$---as it should
in large-momentum transfer reactions---it can actually be omitted in the 
above equation. 
In addition, 
$$\frac{d^2\sigma_S}{dm' \, ^2dt} \propto p'|A_S(s,t,m^2)|^2 \ ,$$
where $\sigma_S$ denotes the partial cross section corresponding to $A_S$ 
-- and where we included again the explicit $S$ dependence as a reminder. 
Thus we can replace $|A(m^2)|^2$ in Eq.
(\ref{almostfinal}) by the cross section, where all constant
prefactors can be omitted again because 
the integral vanishes for a constant argument of the logarithm.
The result after performing these manipulations is given in Eq. (\ref{final}).

Since Eq. \eqref{almostfinal} gives an integral expression not only for the
threshold value of the phase shift but also for the energy dependence, one
might ask whether it is possible to extract, besides the scattering 
length, also the effective range directly from production data. Unfortunately, 
this is not very practical. By taking the derivative with respect to $m^2$ 
of both sides of Eq. \eqref{almostfinal} one sees that one can
get only an integral representation for the product $a^2_S((2/3)a_S-r_S)$ 
but not for the effective range $r_S$ alone.
Thus, although the corresponding integral is even better
behaved than the one for $a_S$ in Eq. \eqref{final} (it has a weaker dependence 
on $m^2_{max}$), the attainable accuracy of $r_S$ will always be limited by 
twice the relative error on $a_S$. 
In fact, 
a more promising strategy to pin down the $YN$ low energy parameters 
might be to
fix the scattering lengths from production reactions and then use the existing
data on elastic scattering to determine the effective range parameters. 
In this case one would benefit from having to do an interpolation 
between the results at threshold, constrained from production reactions,
and the elastic $YN$ data available only at somewhat higher energies.
This should significantly improve the accuracy 
as compared to analyses as, e.g., in Ref. \cite{data1b} which 
rely on elastic scattering data alone.
 
\begin{figure}[t]
\vspace{10cm}
\includegraphics{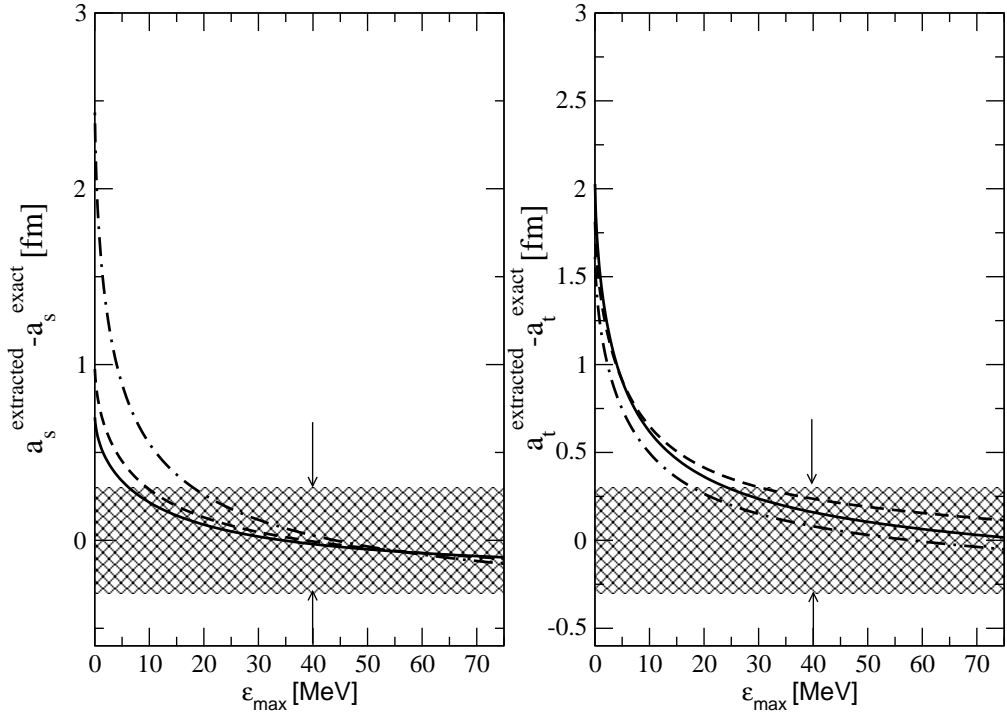}
\caption{Dependence of the extracted scattering lengths
on the value of the upper limit of integration, $\epsilon_{max}$.
Shown is the difference from the exact results for the spin singlet 
scattering length $a_s$ (left panel)
and the spin triplet scattering length $a_t$ (right panel). 
The solid and the dot--dashed line correspond respectively to model 
NSC97a and NSC97f of Ref. \protect{\cite{rijken}} and the dashed one
corresponds to the $YN$ model of Ref. \protect{\cite{jueln}}.
The shaded area indicates the estimated error of the proposed method
and the arrows indicate the value for $\epsilon_{max}$ as estimated based on
scale arguments.
}
\label{cdep} 
\end{figure}

\subsection{Relevant scales and error estimates}

The next step is to estimate an appropriate range of values for
$m_{max}^2$ as well as the systematic uncertainty of the method. Naturally, $m_{max}^2$
has to be large enough to allow resolution of the relevant structures of the
final state interaction. (Note that $m_{max}^2 \to m_0^2$ leads to a vanishing value
for the scattering length, c.f. Eq.  (\ref{final}).)  On the other hand
it should be as small as possible in order to ensure that the $\Lambda N$ system is 
still produced predominantly in an $S$--wave but also, as mentioned, 
to avoid the explicit inclusion of further right-hand cuts such as the one resulting 
from the opening of the $\Sigma N$ channel. 

Eq. (\ref{arform}) suggests that we should choose $m_{max}^2$ large
enough to allow values of $p'$ such that $ap' \sim 1$. Thus, for
non--relativistic kinematics, we find $m_{max}^2 \sim m_0^2(1+1/(\mu
m_0 a^2))$.  A more transparent quantity is $\epsilon_{max}$ defined by  
$$\epsilon_{max} = m_{max}-m_0 \simeq \frac1{2\mu a^2}  \ , $$  
i.e. the maximum kinetic energy of the $\Lambda N$ system for which it 
contributes to the integral of Eq. (\ref{final}). 
For $a\sim 1$ fm we get $\epsilon_{max} \sim$ 40 MeV. 
This is well below the $\Sigma N$ threshold which corresponds to 
an energy of 75 MeV and the threshold for the $\pi \Lambda N$ channel
which is at 140 MeV.

Let us now estimate the uncertainty. 
Except for the neglect of the kaon--baryon interactions, 
Eq. (\ref{almostfinal}) is exact. Therefore $\delta a^{(th)}$ -- the theoretical uncertainty
of the scattering length extracted using Eq. (\ref{final}) -- is given by the integral
\begin{eqnarray}
\nonumber \delta a^{(th)}&=& -\lim_{{m}^2\to
m_0^2}\frac1{2\pi}\left(\frac{m_\Lambda+m_N} {\sqrt{m_\Lambda
m_N}}\right) \\ & & \qquad \times {\bf P}
\int_{m_0^2}^{m_{max}^2}\frac{\log{|\tilde \Phi(m_{max}^2,m' \, ^2)|^2}}
{\sqrt{m' \, ^2-m_0^2}(m' \, ^2-m^2)}\sqrt{\frac{m_{max}^2-m^2}{m_{max}^2-m'
\, ^2}}dm' \, ^2.
\label{deltaath}
\end{eqnarray} 
Since $\log{|\tilde
\Phi(m_{max}^2,m^2)|^2}=\log{|\Phi(m^2)|^2}+\log{|\Phi_{m_{max}^2}(m^2)|^2}$, 
we may write $\delta a^{(th)}=\delta a^{(lhc)}+\delta
a^{m_{max}}$, where the former, determined by $\Phi(m^2)$, is controlled
by the left hand cuts and the latter, determined by $\Phi_{m_{max}^2}(m^2)$,
by the large energy behavior of the $\Lambda N$ scattering
phase shifts. 
The closest left hand singularity is that introduced by the exchange of light
mesons ($\pi$ and $K$) in the production
operator and it is governed by the momentum transfer. 
Up to an
irrelevant overall constant, we may therefore estimate the variation
 of $\Phi \sim 1+\kappa(p'/p)^2$, where we assume $\kappa$ to be of
 the order of 1. Evaluation of the integral (\ref{deltaath}) then
 gives 
 $$\delta a^{(lhc)} \sim \kappa(p'_{max}/p^2)\sim 0.05 \ \mbox{fm} \ . $$
Here we used $p'_{max}\, ^2=2\mu \epsilon_{max}$ with $\epsilon_{max} \sim
40$ MeV and $p \sim 900$ MeV, where the latter is the center-of-mass momentum 
in the initial $NN$ or $\gamma d$ state that corresponds to the 
$K\Lambda N$ threshold energy.
Concerning $\delta a^{m_{max}}$ we start from the definition of 
$\Phi_{m_{max}^2}(m^2)$ in Eq. (\ref{dis3}) from which one easily derives
\begin{eqnarray}
 |\delta
 a^{m_{max}}|=\frac2{\pi p'_{max}}\left|\int_0^\infty
 \frac{\delta(y)dy}{(1+y^2)^{(3/2)}}\right| \leq \frac2{\pi p'_{max}}
 |\delta_{max}| \ , \label{dam} 
\end{eqnarray}
where $y^2=(m^2-m_{max}^2)/(m_{max}^2-m_0^2)$. Thus, in order to obtain 
an estimate for $\delta a^{m_{max}}$, we need to make an
assumption about the maximum value of the elastic $\Lambda N$ phase
for $m^2 \ge m_{max}^2$. 
In addition, it is important to note that the denominator in the integral 
appearing in Eq. (\ref{dam}) strongly suppresses large values of
$y$. Since for none of the considered $YN$ models 
\cite{maessen,juel,juel1,rijken,jueln} does
$\delta_{max}$ exceed 0.4 rad, we arrive at the estimate $\delta
a^{m_{max}}\sim 0.2$ fm. 
Note that when using the phase shifts as given by those
models directly in the integral, the value for $\delta a^{m_{max}}$ is
significantly smaller, since for all models considered the phase changes sign 
at energies above $m_{max}^2$.  Combining the two error estimates, we conclude
$$
\delta a \lesssim 0.3 \ \mbox{fm} \ .
$$

Another source of uncertainty comes from possible final state interactions in
the $K N$ and $K\Lambda$ systems. Their effects have been neglected in our
considerations so far for a good reason: their influence is to be energy
dependent and therefore no general estimate for the error induced by them can
be given. For example, baryon resonances are expected to have a significant
influence on the production cross sections \cite{shyam}.
However, it is possible to examine the influence of the meson--baryon 
interactions experimentally, namely through a Dalitz plot analysis.
This allows to see directly, if the area where the $\Lambda N$ interaction is
dominant is isolated or is overlapping with resonance structures
\cite{report}. In addition, to quantify the possible influence of the meson--baryon
interactions on the resulting scattering lengths, the extraction procedure proposed
above is to be performed at two or more different beam energies. If the obtained
scattering lengths agree, we can take this as a verification that the result
is not distorted by the interactions in the other subsystems because their
influence necessarily depends on the total energy, whereas the extraction
procedure does not. 

\section{Test of the method: comparison to model calculations}

The most obvious way to test Eq. \eqref{final} would be to apply it to
reactions where the involved scattering lengths are known as it is the case for 
any large momentum transfer reaction with a two nucleon final state. 
Unfortunately, as far as we know there are is no experimental information
on $nn$ or $pn$ mass spectra with sufficient resolution to allow the application 
of Eq. \eqref{final}. 
For existing invariant mass spectra with a $pp$ final state, as
 given, e.g., in Ref. \cite{bilger} for the reaction $pp\to pp\pi^0$, our formula
 is not applicable due to the presence of the Coulomb interaction 
 that strongly distorts the invariant mass spectrum exactly in
 the regime of interest. Moreover, one has to keep in mind that the authors of 
 Ref. \cite{bilger} were forced to assume already a particular $m^2$ dependence of 
 the outgoing $pp$ system at small values of $m^2$ for their analysis because of the 
 limited detector acceptance. Thus, a test with those data would not be
conclusive, anyway. 
 
 Therefore, we chose a different strategy in order to test our method.
 We apply Eq. \eqref{final} to the results of a microscopic model for the
 reaction $pp\to K^+p\Lambda$ \cite{model}. (We expect that application to the
 reaction $\gamma d\to K^+n\Lambda$ gives similar results.) In this model the
 $K$ production is described by the so-called $\pi$ and $K$ exchange
 mechanisms. Thereby a meson ($\pi$ or $K$) is first produced from one of the
 nucleons and then rescattered on the other nucleon before it is emitted (cf.
 Ref. \cite{model} for details). The rescattering amplitudes ($\pi N \to KY$,
 $KN \to KN$) are parameterized by their on-shell values at threshold. The $YN$
 final state interaction, however, is included without approximation. 
 Specifically, the coupling of the $\Lambda N$ to the $\Sigma N$ system is 
 treated with its full complexity. For the present test calculations we used all
 the $YN$ models of Ref. \cite{rijken} as well as the $YN$ model of Ref.
 \cite{jueln}. Thus, the test calculation will allow us to see a possible
 influence of the $\Sigma N$ channel as well as that of the left hand
 singularities induced by the production operator. In addition, we can study
 the effect of the value of the upper integration limit $m_{max}^2$ or,
 equivalently, $\epsilon_{max}$ on the scattering length extracted.

We now proceed as follows: we calculate the invariant mass spectra
needed as input for Eq. (\ref{final}), utilizing the $YN$ models mentioned 
above. Then we use that equation to extract the scattering lengths 
from the invariant mass plots.
In Fig. \ref{cdep} the difference between the extracted scattering length
and the exact one is shown as a function of $\epsilon_{max}$ for several $YN$ 
models. The direction of kaon emission was chosen as 90 degees in the center of
mass. Considered are the two ``extreme'' models of Ref. \cite{rijken}, namely
NSC97a, with singlet scattering length $a_s=-0.71$ fm and
triplet scattering length $a_t=-2.18$ fm, and NSC97f ($a_s=-2.51$ fm and
$a_t=-1.75$ fm), and the new J\"ulich model ($a_s=-1.02$ fm and
$a_t=-1.89$ fm). The last was included here because it 
should have a significantly different short range behavior compared
to the other two. As can be seen from the figure, for
$\epsilon_{max} \simeq$ 40 MeV in all cases 
the extracted value for the scattering length does not deviate
from the exact one by more than the previously estimated 0.3 fm, as is
indicted by the shaded area in Fig. \ref{cdep}.
This is in accordance with the error estimates given in the previous section, where 
$\epsilon_{max}\sim 40$ MeV -- in the figure highlighted by the arrows -- was deduced 
from quite general arguments.
In addition, this investigation also suggests that the $\Sigma N$ cuts, included in the 
model without approximation, do not play a significant role in the determination of the
scattering length.

\begin{figure}[t]
\vspace{10cm}
\includegraphics{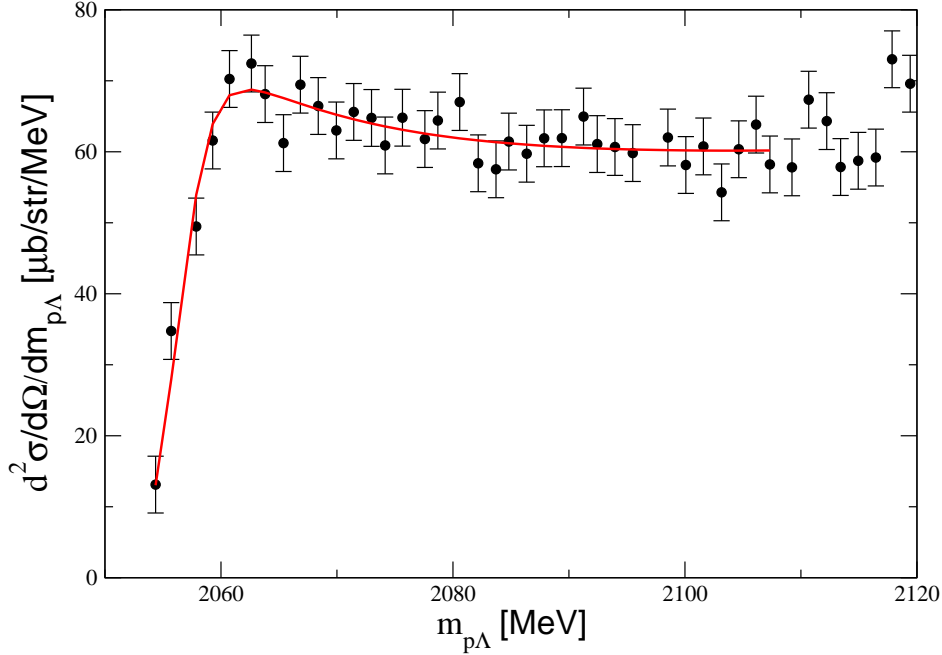}
\caption{$pp \to K^+X$ inclusive missing mass spectrum at 
$T_p$ = 2.3 Gev for kaon laboratory scattering angles
$\theta^L_K$ = 10$\pm 2^o$ \protect{\cite{saclay}}. The
solid line is our fit as described in the text. 
Note that the scattering length extracted is insensitive to the 
parameterization of the data used (see Appendix for details).}
\label{datfig} 
\end{figure}

\section{Application to existing data}

 As was stressed in the derivation, Eq. (\ref{final}) can
be applied only to observables (cross sections) that are dominated 
by a single partial wave. This is indeed the case for specific polarization
observables for near-threshold kinematics as we show in the
appendix \ref{obs} for the reaction $pp\to NK\Lambda$. 
Unfortunately, at present no suitable polarization data are available that 
could be used for a practical application of our formalism. 
Thus, in order to demonstrate the potential of the proposed method,
we will apply it to the unpolarized data for the reaction $pp\to K^+X$ 
measured at the SPES4 facility in Saclay \cite{saclay}.
Obviously, since we only consider the small invariant mass
tail of the double differential cross section, we automatically
project on the $p\Lambda$ $S$--waves. 
The data, however, still contains
contributions from both possible final spin states -- spin triplet as well
as spin singlet. Thus the scattering length extracted from those data
is an effective value that averages over the spin-dependence of
the $p\Lambda$ interaction in the final state and also over the
spin dependence of the production mechanism with unknown weightings.
It does not allow any concrete conclusions on the elementary
$p\Lambda$ ($^1S_0$ or $^3S_1$) scattering lengths. We should mention,
however, that some model calculations of the reaction $pp \to K^+p\Lambda$
\cite{model,laget} indicate that the production
mechanism could be dominated by the spin-triplet configuration. If this
is indeed the case then the production process would act like a
spin filter and the scattering length extracted from the
Saclay data could then be close to the one for the $p\Lambda$ $^3S_1$
partial wave.

For our exemplary calculation we use specifically the data on the 
$pp \to K^+X$ inclusive missing mass spectrum at 
$T_p$ = 2.3 GeV for kaon laboratory scattering angles
$\theta^L_K$ = 10$\pm 2^o$\footnote{For the given energy and
$\epsilon_{max}$=40 MeV a fixed kaon angle is almost
equivalent to a fixed $t$ as is required for the dispersion integral.} 
\protect{\cite{saclay}} (cf. 
Fig. \ref{datfig}). For convenience we represent those data
in terms of simple analytical functions (the result of the best fit
is shown as the solid line in the figure) as described in appendix \ref{fit}. 
Please note, however, that the advantage of the proposed method is that
the scattering length extracted is independent of the particular
analytical parameterization employed; one can even use the data directly.
In particular we do not need to assume that the elastic $\Lambda N$ interaction can be
represented by the first two terms of an effective range expansion.
The analytical representation of the data is then used to evaluate the 
corresponding scattering length utilizing Eq. (\ref{final}). 
Details of the fitting as well as how we extract the uncertainty
in the scattering length can be found in the appendix.

We find a scattering length of $(-1.5 \pm 0.15 \pm 0.3)$ fm, 
where the first error corresponds to the uncertainty from the data and the
second one is the theoretical uncertainty discussed previously.
Thus, already in the case of an experimental resolution of about 2 MeV the
experimental error is smaller than the theoretical one.

\section{Summary}

In this paper we presented a formalism
that allows one to relate spectra from large-momentum transfer reactions,
such as $pp\to K^+p\Lambda$ or $\gamma d\to K^+n\Lambda$, directly to 
the scattering length of the interaction of the final state particles.
We estimated the theoretical error of the analysis to be less than
0.3 fm. This estimate was confirmed by comparing results obtained with
the proposed formalism to those of microscopic model calculations for the 
specific reaction $pp\to K^+\Lambda p$.

The formalism can 
be applied only to observables (cross sections) that are dominated 
by a single partial wave. This requirement is fulfilled by specific 
polarization observables for $\Lambda N$ invariant masses near 
threshold as we show in the
appendix \ref{obs} for the reaction $pp\to NK\Lambda$. Corresponding
experiments require a polarized beam or target in order to pin down the
spin-triplet scattering length and polarized beam and target for
determining also the spin-singlet scattering length.

Since at present no suitable polarization data are available 
we demonstrated the potential of the proposed method by applying 
it to unpolarized data from the reaction $pp\to K^+X$ 
measured at the SPES4 facility in Saclay \cite{saclay}.
Thereby a scattering length of $(-1.5 \pm 0.15 \pm 0.3)$ fm was 
extracted. This more academic exercise demonstrates that the resolution 
of 2 MeV of the SPES4 spectrometer is already sufficient 
to obtain scattering lengths with an experimental error of less than
0.2 fm. Such an accuracy would be already an improvement over  
the present situation. 
  
The method of extraction of the $YN$ scattering lengths discussed
here should be viewed as an alternative method to proposed analyses of
other reaction channels involving the $YN$ system like
$K^-d\to \gamma n\Lambda$. The assumptions that go into the analyses 
are very different and therefore carrying out both analyses is useful in 
order to explore the systematic uncertainties.

{\bf Acknowledgments}

We thank A. Kudryavtsev for stimulating discussion and J. Durso
for critical comments and a careful reading of the manuscript. 
We are grateful to 
M. P. Rekalo and E. Tomasi-Gustafsson for inspiring and educating discussions
about spin observables.

\appendix

\section{Method of the Analysis}
\label{fit}

In order to evaluate the integral in Eq. \eqref{final} for the 
experimental results of Ref. \cite{saclay} we first fit the data 
utilizing the following parameterization
for the amplitude squared (c.f. Eq. (\ref{almostfinal})): 
\begin{eqnarray}
|A(m)|^2=\exp{\left[C_0+\frac{C_1^2}{(m^2-C_2^2)}  \right]}.
\label{param1}
\end{eqnarray} 
The advantage of such a parameterization is that the integral in 
Eq. \eqref{final} can be calculated analytically and one gets for the 
scattering length 
\begin{eqnarray}
a(C_1,C_2)=-\frac12C_1^2\sqrt{\left(\frac{m_0^2}{m_Nm_\Lambda}\right)\frac{(m_{max}^2-m_0^2)}{(m_{max}^2-C_2^2)(m_0^2-C_2^2)^3}}.
\label{aofc}
\end{eqnarray}
One can extend Eq. \eqref{param1} easily by including in the exponent 
additional terms such as $C_3^2/(m^2-C_4^2)$ if required for a satisfactory
analytical representation of the data. 
The accuracy of the presently available experimental results, however, 
did not necessitate such an extension, for already the form given 
in Eq. \eqref{param1} ensured a $\chi^2$ per degree of freedom of less than 1.

In Ref. \cite{hinterberger} it is stressed that there is a calibration
uncertainty in the data of Ref. \cite{saclay}. The actual value of this
uncertainty was determined by including as a free parameter a shift $\Delta m$ in 
the fitting procedure. In this way a value of $\Delta  m=1.17$ MeV was found.
We use the same value. In addition, the finite resolution of the
detector $\sigma_m=2$ MeV has to be taken into account. Thus the function 
\begin{eqnarray}
\frac{d^2 \sigma(m)}{dmd\Omega_K}=
\int_{-\infty}^\infty\left(\frac{d^2
  \sigma(m')}{dm'd\Omega_K}\right)_{0}\frac1{\sqrt{2\pi}\sigma_m}
\exp{\left[\frac{-(m-m'-\Delta  m)^2}{\sigma_m}\right]}dm' \ ,
\label{smeared}
\end{eqnarray} is to be compared to the data,
where $(d^2
  \sigma(m')/(dm'd\Omega_K))_{0}$ is calculated from the amplitude
Eq. \eqref{param1}.

The probability for the data to occur under the assumption that the
cross section given in Eq. \eqref{smeared} is true is given by the Likelihood
function. Here we assume the data distribution to be Gaussian: 
\begin{eqnarray}
\mathcal{L}(\{\mbox{data}\}|C0,C1,C2;I) \propto
\exp{\left[-\frac12\chi^2(C_0,C_1,C_2) \right]} \ ,
\label{llh}
\end{eqnarray}
where the standard $\chi^2$ function appears in the exponent. The letter $I$
that appears in the argument of $\mathcal{L}$ is meant to remind the reader that
some assumptions had to be made in order to write down Eq. \eqref{llh}.

The distribution of interest to us is the probability distribution of the
scattering length $\mathcal{L}(a|\{\mbox{data}\};I)$, given the data, that can
be written as \cite{sivia}
\begin{eqnarray}
\mathcal{L}(a|\{\mbox{data}\};I)&=& \nonumber \\ 
&\mathcal{N}&\int \prod_idC_i \mathcal{L}(\{\mbox{data}\}|C0,C1,C2;I)
\delta(a-a(C_1,C_2)) \ ,  
\label{pdf}
\end{eqnarray}
where $\mathcal{N}$ is a normalization constant to be fixed through $\int
\mathcal{L}(a)da=1$ and $a(C_1,C_2)$ is given in Eq. \eqref{aofc}.

It turns out that the Likelihood function of Eq. \eqref{llh} is strongly
peaked. For the numerical evaluation of Eq. \eqref{pdf} we therefore linearize
the individual terms with respect to the parameters $C_i$.
Within this linear approximation the
resulting probability density function $\mathcal{L}(a)$  also takes a
Gaussian form.
To test that there is no significant dependence of the result on the 
parameterization used for the data we also used the Jost function in the
analysis. The results for the scattering lengths determined with the two
different methods agreed within the statistical uncertainty.

\section{Spin dependence of the reaction $pp\to \Lambda pK^+$}
\label{obs}

The aim of this appendix is to show how 
polarization measurements for the reaction $pp\to \Lambda pK^+$ can be
used to disentangle the spin dependence of the production cross section.

In terms of the so called Cartesian polarization observables,
the spin--dependent cross section can be written as \cite{meyerpol}
\begin{eqnarray}
\nonumber
\sigma (\xi, \vec P_b, \vec P_t, \vec P_f)
&=& \sigma_0(\xi)\left[1+\sum_i ((P_b)_iA_{i0}(\xi)+(P_f)_iD_{0i}(\xi))\right. \\
& &\left.\phantom{\sigma_0(\xi)[1} \ +\sum_{ij}(P_b)_i(P_t)_jA_{ij}(\xi)+ ...\right]
\label{obsdef}
\end{eqnarray}
where $\sigma_0(\xi)$ is the unpolarized differential cross section, the
labels $i,j$ and $k$ can be either $x,y$ or $z$, and
$P_b$, $P_t$ and $P_f$ denote the polarization vector of beam, target and one
of the final state particles, respectively.
All quantities are functions
of the 5 dimensional phase space, abbreviated as $\xi$.
The observables shown explicitly in Eq. (\ref{obsdef}) include the beam
analysing powers $A_{i0}$, the corresponding quantities for the final state
polarization $D_{0i}$, and the spin correlation coefficients $A_{ij}$.
Polarization observables not relevant for this work are not shown explicitly.
All those observables that can be defined by just
exchanging $\vec P_b$ and $\vec P_t$, such as the target analyzing power $A_{0i}$,
are not shown explicitly.

The most general form of the transition matrix element may be written
as \cite{report}
\begin{equation}
{\mathcal M}=H({\mathcal I}\, {\mathcal I \, '})+ i\vec Q \cdot 
(\vec S \, {\mathcal I '})+i\vec A \cdot (\vec S \, '\, {\mathcal I})
 +(S_i \, S_j \, ')
B_{ij} \ ,
\end{equation}
where $\vec S=\left(\chi_2^T\sigma_y\vec \sigma \chi_1\right)$ and
$\vec S \, '=\left(\chi_4^\dagger \vec \sigma \sigma_y 
(\chi_3^T)^\dagger\right)$ are to be used for spin triplet initial and final
states, respectively, and 
${\mathcal I}=\left(\chi_2^T\sigma_y \chi_1\right)$ and
${\mathcal I \, '}=\left(\chi_4^\dagger \sigma_y (\chi_3^T)^\dagger\right)$
are to be used for the corresponding spin singlet states.
Here it is assumed that the outgoing baryons are non
relativistic, as is the case for the kinematics considered.

\begin{figure}[t]
\vspace{10cm}
\includegraphics{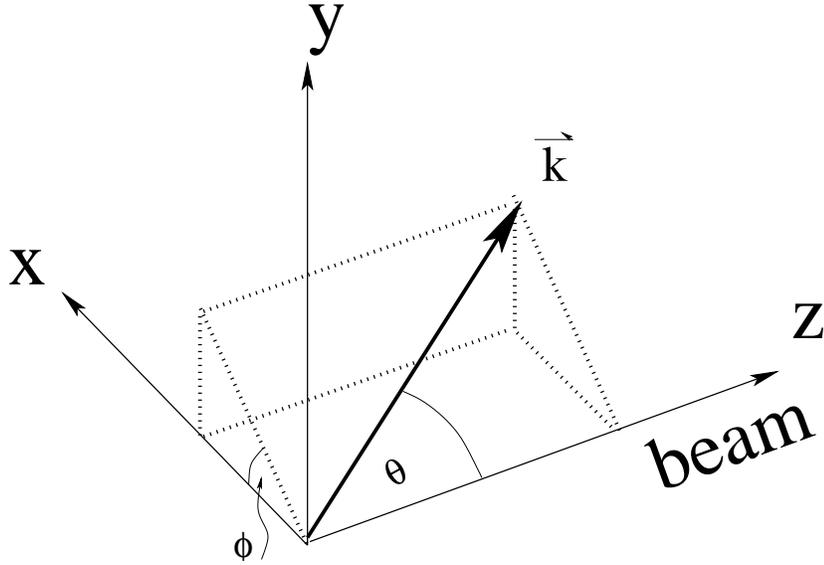}
\caption{The kaon momentum $\vec k$ plotted in the coordinate system used;
 the $z$--axis is along the beam direction.}
\label{koorsys} 
\end{figure}

Given this form it is straight forward to evaluate the expression for the
various observables. E.g.
\begin{eqnarray}
4\sigma_0 &=& |H|^2+|Q_m|^2+|A_m|^2+|B_{mn}|^2 \, , \label{si0def}
 \\
4A_{0i}\sigma_0 &=&
i\epsilon_{ikn}\left( Q_k^*Q_n+B_{km}^*B_{nm}\right)+2\mbox{Im}
\left(B_{im}^*A_m-Q_i^*H\right) \, , \label{a0idef} \\
4D_{0i}\sigma_0 &=&
-i\epsilon_{ikn}\left( A_k^*A_n+B_{mk}^*B_{mn}\right)-2\mbox{Im}
\left(B_{mi}^*Q_m-A_i^*H\right) \, , \label{d0idef} \\
4A_{ii}\sigma_0 &=&
-|H|^2+|Q_m|^2-|A_m|^2+|B_{mn}|^2
-2|Q_i|^2-2|B_{im}|^2 \, , \label{aiidef}
\end{eqnarray}
where the indices $m$, $k$ and $n$ are to be summed.
By definition, the spin triplet final state amplitudes contribute to $A$
and $B$ only, whereas the spin singlet amplitudes contribute to $H$
and $Q$. By definition, the spin triplet states interfere with the spin singlet
states only, if the final state polarization is measured. In the examples
given above this is the case for $D_{0i}$ only. This simplifies the
analysis considerably.
 Eq. (\ref{final}) needs as input only those cross sections
that have low relative energies, where we can safely assume the
$\Lambda N$ system to be in an $S$--wave. Thus, in order to construct
the most general transition amplitude under this condition, we need to
express the quantities $H$, $\vec Q$, $\vec A$ as well as $B_{ij}$ in
terms of the vectors $\vec k$ (the momentum of the kaon) and $\vec p$
(the momentum of the initial protons)---c.f. Fig. \ref{koorsys}.  The Pauli Principle
demands for a proton--proton initial state, that $\vec S$ appears with
odd powers of $\vec p$ and terms that do not contain $\vec S$ appear
with even powers of $\vec p$. Correspondingly, due to parity conservation
these amplitudes need to be even and odd in $\vec k$ respectively,
since the kaon has a negative intrinsic parity.
We may therefore write
\be
\nonumber
H &=& 0 \, , \\
\nonumber
\vec Q &=& s_1(\vec k \cdot \hat p)\vec k  + s_2\hat p \, , \\
\nonumber
\vec A &=& t_1\hat p (\hat p\cdot \vec k)+t_2\vec k \, , \\
B_{ij} &=& \epsilon_{ijk}\left(t_3 k_k (\vec k\cdot \hat p)+t_4 \hat p_k\right) 
 +(\vec k \times \hat p )_i\left(t_5k_j+t_6(\hat p\cdot \vec k) \hat p_j\right) \, ,
\label{HQAB}
\ee
where $\hat p = \vec p/|\vec p|$. The amplitudes $s_i$ and $t_i$ appearing 
in the above equations are functions of $p^2$, $k^2$ and 
$(\vec k \cdot \vec p)^2$. Now we want to identify observables that either 
depend only on (some of) the $s_i$ and thus are sensitive to
the spin singlet $\Lambda N$ interaction only or on (some of) the $t_i$ and 
thus are sensitive to the spin triplet $\Lambda N$ interaction only.
It should be stressed in this context that it is entirely due to the 
relation $H=0$ -- a direct consequence of the selection rules -- 
that one can disentangle the spin triplet and the spin singlet channel 
without making assumptions about the kaon partial wave, 
as was first observed in Ref. \cite{vigdor}.


Using Eqs. (\ref{HQAB}) 
all observables can be expressed in terms of the
amplitudes $s_i$ and $t_i$, where the former correspond to a spin singlet
$YN$ pair in the final state, while the latter correspond to a
spin triplet pair. We find
\begin{eqnarray}
A_{0i}\sigma_0 = \frac12(\vec k\times \hat p)_i
\left(\alpha+\beta (\vec k\cdot \hat p)+\gamma (\vec k\cdot \hat p)^2\right)
\end{eqnarray}
where
\begin{eqnarray}
\alpha &=& \mbox{Im}(t_4^*t_2+t_5^*t_2\vec k \, ^2) \ ,\\
\beta &=& \mbox{Im}(t_4^*t_3+s_2^*s_1) \ ,\\
\gamma &=& \mbox{Im}(t_5^*t_1+t_6^*t_1+t_6^*t_2-t_3^*t_1) \ .
\end{eqnarray}
We may
 express the results in the coordinate
system defined by the beam along the $z$ axis and the 
$x$ and $y$ directions through the polarizations (c.f. Fig. \ref{koorsys}). Then
we find for the analysing power
\begin{eqnarray}
\nonumber
A_{0y}\sigma_0
&=& -\frac{1}{4}k^2\beta\sin (2\theta )\cos (\phi) \\
& & \ \ \ \ \ +\sin (\theta )\cos (\phi)\mbox{(spin
triplet only)} \ ,
\end{eqnarray}
where we used $(\hat p \times \vec k)_y=k\sin (\theta) \cos (\phi)$ and
$\hat p\cdot \vec k = k \cos (\theta)$.
Thus, the contributions from all those partial waves where
the final $\Lambda N$ system is in an $^1S_0$ state
vanish when the kaon goes out
in the $xy$--plain. In addition, the term proportional to $\beta$ is the only
one that is odd in $\cos (\theta)$, and thus any integration with respect to
the angle $\theta$ of $A_{0y}\sigma_0$ symmetric around $\pi/2$ removes any
spin singlet contribution from the observable.
Please note that for any given energy it should be checked whether
the range of angular integration performed is consistent with the requirement
that the reaction was dominated by a large momentum transfer -- kaon emission at
90 degrees maximizes the momentum transfer and thus minimizes the error in the
extraction method.

Analogously one can show for the combination $(1-A_{xx})\sigma_0$ that
\be
\nonumber
(1-A_{xx})\sigma_0
&=& \frac12\left(|\vec A|^2+|Q_x|^2+|B_{xm}|^2\right) 
\\
&=& 
\frac18k^4|s_1|^2\sin ^2(2\theta )\cos ^2(\phi) 
+\mbox{(spin
triplet only)} \ ,  
\ee
i.e. that it 
also allows to isolate the amplitudes with the $YN$ system in the
spin triplet. Furthermore, 
\be
\nonumber
(1+A_{xx}+A_{yy}-A_{zz})\sigma_0
\\ \nonumber & & 
\!\!\!\!\!\!\!\!\!\!\!\!\!\!\!\!\!\!\!\!\!\!\!\!\!\!\!\!\!\!\!\!\! =
|Q_z|^2+|B_{zm}|^2 \\
& & 
\!\!\!\!\!\!\!\!\!\!\!\!\!\!\!\!\!\!\!\!\!\!\!\!\!\!\!\!\!\!\!\!\! =
\frac14k^4|t_3|^4\sin ^2(2\theta )
+\mbox{(spin
singlet only)} \ 
\label{axxfinal}
\end{eqnarray}
and therefore allows to isolate the spin singlet amplitudes. 
(Note that in both cases a summation over $m$ is to be performed.)
Thus, for all those partial waves where the
final $YN$ system is in the $^1S_0$ final state, $(1-A_{xx})$
vanishes, when the kaon goes out in the $yz$-- or in the $xy$--plain and for
all those partial waves where the final $YN$ system is in the
$^3S_1$ final state, $(1+A_{xx}+A_{yy}-A_{zz})$ vanishes, when the kaon goes
out in the $xy$--plain or in forward direction. These results hold
independent of the orbital angular momentum of the kaon.

So far we assumed that higher partial waves of the $\Lambda N$
system do not play a role -- since we restricted ourselves to small relative 
energies. However, it is also possible to check this experimentally.
First of all the angular distribution
of the $YN$ system $d\sigma /d\Omega_{p'}$, where
$\vec p \, '$ is the relative momentum of the two outgoing 
baryons, needs to be flat. However, in the presence of spins even
flat angular distributions can stem from higher partial waves \cite{kanzonew}.
To exclude this possibility the $\Lambda$ polarization can be used.
For this test we need an observable that vanishes in the
absence of higher $\Lambda N$ partial waves. The easiest choice
here is $D_{0i}$. It is straight forward to show that
$$
D_{0i} \propto (\vec p \times \vec k)_i \ ,
$$
as long
as there are only $S$ waves in the $\Lambda N$ system.
Therefore, in the absence of higher partial waves in the $YN$
system $D_{0z}$ has to vanish.

\end{document}